\documentclass[aps,prb,twocolumn,groupedaddress,amsmath,amsfonts,amssymb]{revtex4-1}
\usepackage[english]{babel}
\usepackage{ucs}
\usepackage[utf8x]{inputenc}
\usepackage{graphicx}
\usepackage{bm}
\usepackage{bbm}
\usepackage{empheq}
\usepackage{mathbbol}

\begin{document}

\title{All-optical control of the spin state in the $NV^-$-center in diamond}

\author{Florian Hilser and Guido Burkard}
\affiliation{Department of Physics, University of Konstanz, D-78457 Konstanz, Germany}

\begin{abstract}
We describe an all-optical scheme for spin
manipulation in the ground-state triplet of the negatively charged 
nitrogen-vacancy (NV) center in diamond.
Virtual optical excitation from the $^3A_2$ ground
state into the $^3E$ excited state allows for spin rotations by virtue of
the spin-spin interaction in the two-fold orbitally degenerate excited state.
We derive an effective Hamiltonian for optically induced 
spin-flip transitions within the ground state spin triplet due to
off-resonant optical pumping.
Furthermore, we investigate the spin qubit formed by the Zeeman
sub-levels with spin projection $m_S = 0$ and $m_S = -1$ along the NV
axis around the ground state level anticrossing with regard to full optical control of 
the electron spin. 
\end{abstract}

\maketitle

\section{Introduction}

Nitrogen-vacancy (NV) centers in diamond have attracted much attention
in research related to quantum computation \cite{Hanson2008}
due to their key advantages, such as high stability and 
long spin coherence times \cite{Jel2, Ken,Hanson2006} up to room temperature
and beyond \cite{Toyli2012}. 
The spin coherence time can be increased further by isotopic
engineering \cite{Bal} since only the $^{13}C$ carbon atoms have non-zero nuclear spin, thus contributing to spin decoherence due to hyperfine coupling.
Under resonant optical excitation the $NV^-$ center exhibits a
strong and highly stable zero phonon line at $1.945$ eV \cite{Dav} with an excited 
state lifetime of about $12 \,{\rm ns}$ \cite{Gru}. 
Electron spin resonance analysis of the center has shown that both ground state and excited state are 
spin triplets, which implies that there is an even number of active electrons involved. 
The ground state levels with spin projection $m_S = 0$ and $m_S = -1$
along the NV axis become degenerate in a magnetic field of about $1025$ G.
Optical pumping causes a spin polarization of the ground state
\cite{Loub, Har, Har2} 
that can be attributed to a spin-orbit induced intersystem crossing 
with an intermediate singlet state\cite{Man}.
When the zero field splitting is larger than the optical linewidth, repeated optical excitation leads to a 
spin selective steady state population in the lowest $m_S = 0$ level of the ground state, generating a non-Boltzmann steady state spin alignment and mixing of spin 
states \cite{He}, so the spin of the ground state can be both initialized and read out optically \cite{San}.

The standard procedure for spin manipulation in the ground state
triplet involves an oscillatory (radio-frequency) magnetic field that
gives rise to electron spin resonance.  In this paper, we describe an
alternative method for full spin control without rf-fields, based
entirely on optical transitions.
All-optical spin manipulation of NV centers could allow for fast
operations with high spatial resolution.
In semiconductor quantum dots, picosecond optical control of 
single electron spins has been achieved
\cite{Mikkelsen2007,Berezovsky2008}.
Optically induced spin rotations in a single NV center in diamond have been 
demonstrated using off-resonant laser excitation \cite{Buckley2010}.  
This type of spin control
relies on the optical Stark effect, i.e., the shift in energy levels
induced by an applied optical field. 
Here, we describe an extension of this scheme which also allows for 
transitions between the three ground state levels, similar to existing 
schemes for coherent population trapping \cite{Togan2011}.
\begin{figure}
 \includegraphics[width=0.52\textwidth]{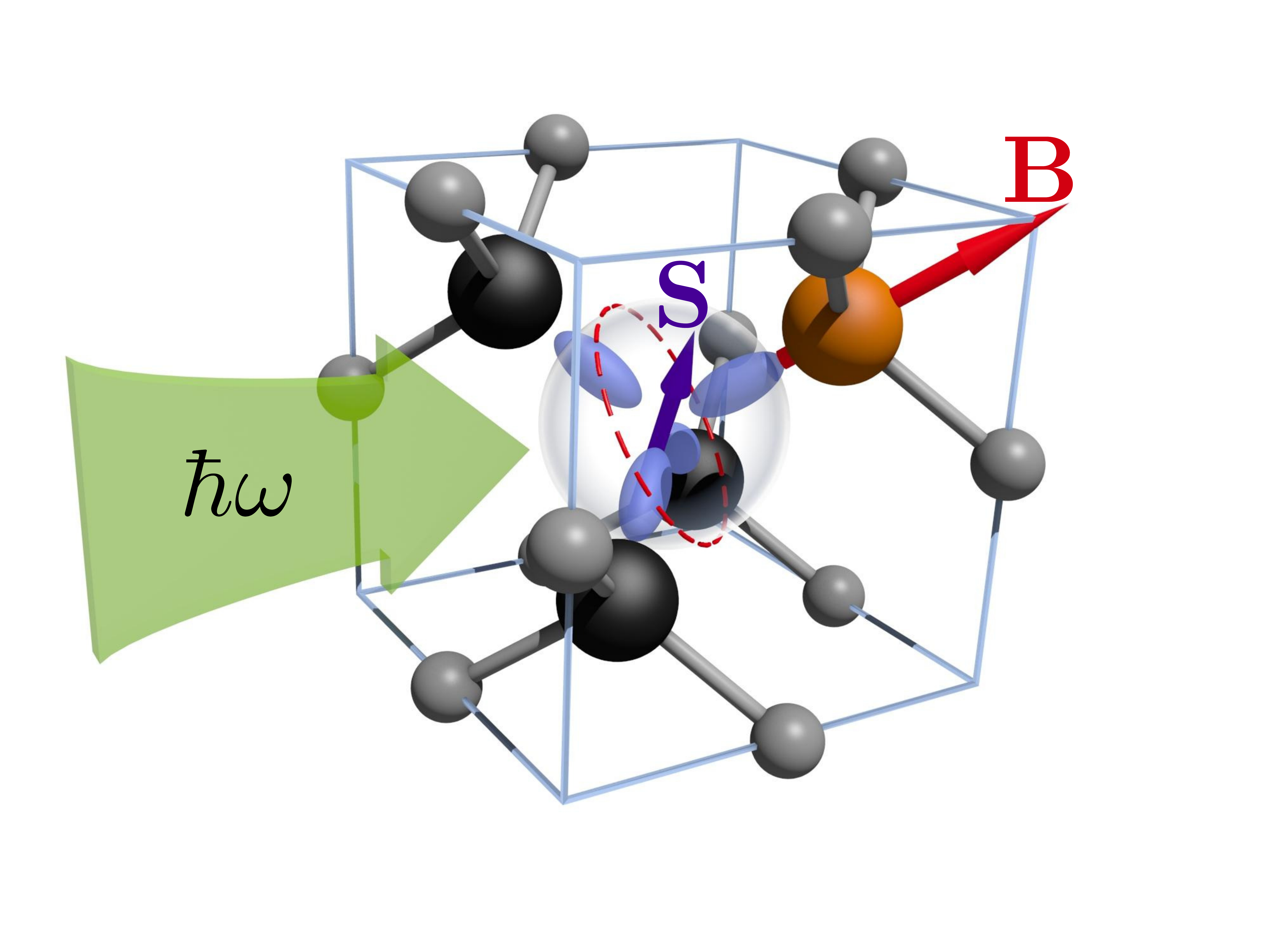}
\caption{NV-center in diamond with equivalent sp$^3$-hybridized dangling
  bonds (blue) $\sigma_i$ ($i=1,2,3$) from the surrounding carbon
  atoms (black) and $\sigma_N$ from the substitutional
  nitrogen atom (orange) overlapping within the vacancy (center). 
The $NV$-axis (long red arrow) defines the $z$-axis of the coordinate
system and, due to the zero field splitting determines
the quantization axis of the electron spin.
The magnetic field $B$ is applied along the  NV-axis.
\label{NV}}
\end{figure}

\section{Model}

To model optical spin rotations in an individual NV-center in diamond,
we start from the commonly used description that fundamentally involves a total
number of six electrons, but can be reduced to an effective
two-electron \cite{Len,Man,Doherty} or, equivalently, two-hole model \cite{Maze}.
The four relevant single-electron orbitals $a_1$, $a_2$, $e_x$, and $e_y$ can
be obtained by projecting the $sp^3$-hybridized dangling bonds $\sigma_{1,2,3}$ of the carbon
atoms and $\sigma_N$ of the nitrogen atom (see Fig.~\ref{NV}) onto the irreducible representations 
of the $C_{3v}$ symmetry group of the NV center \cite{Maze}.
The electron configurations for the ground and first excited state are
obtained as follows:  In the ground state  ($e^2$) configuration, the
lower-energy $a_{1,2}$ orbitals are completely filled with two electrons each, while the
$e_x$ and $e_y$ contain one electron each. The two-fold degenerate excited state
configuration ($ae$) is obtained by promoting another electron from $a_1$ to $e_x$ or $e_y$.
Due to the Coulomb interaction between the two electrons, the spin
triplet lies lowest in energy and forms the ground state $e^2(T)$,
transforming according to the representation $A_2$ of  $C_{3v}$ \cite{Maze}.  
Electric dipole transitions  connect this triplet to the excited state
spin triplet $ea(T)$, transforming according to the $E$
representation.  
The two-fold orbital and three-fold spin degeneracies give rise to a total of six states in
$ea(T)$, compared to three states in  $e^2(T)$.
The spin singlet states will not be of direct importance for our
discussion, and are left out of our model.
The entire state space for our model is thus nine-dimensional.

\subsection{Ground state}

The Hamiltonian of the ground state spin triplet 
in the basis $\left\{^3A_{2-},^3A_{20},^3A_{2+}\right\}
=\left\{|-1\rangle,|0\rangle,|+1\rangle\right\}$
is 
\begin{equation}
H_{\rm gs} = 
\begin{pmatrix}
D_{\rm gs} - g_{\rm gs} \mu_B B & 0 & 0
\\
0 & 0 & 0 
\\ 
0 & 0 & D_{\rm gs} + g_{\rm gs}\mu_B B 
\end{pmatrix},
\end{equation}
where $B$ denotes an external magnetic field aligned with the
$NV$-axis, $g_{\rm gs}$ the Land\'e g-factor, and $\mu_B$
the Bohr magneton.
Around the ground-state level anticrossing (LAC), we can 
split the Zeeman energy into a term that compensates the ground state zero 
field splitting and an additional variation, 
$g\mu_B B = D_{\rm gs} + g\mu_B\delta B$.
The zero field splitting $D_{\rm gs} = 2.88$ GHz \cite{Oort} is caused by 
the reduction of the symmetry in spin space to $C_{3v}$ due to the crystal field.
The absence of orbital degeneracy in the ground state triplet implies 
that strain and spin-orbit interaction have very little effect on the
ground state.

Here, we have neglected the effect of the hyperfine coupling to the 
nuclear spins of the intrinsic nitrogen atom and surrounding $^{13}$C
atoms.  If necessary, the nuclear spin state could be prepared optically.

\subsection{Excited state}
\begin{figure}
\includegraphics[width = 0.48\textwidth]{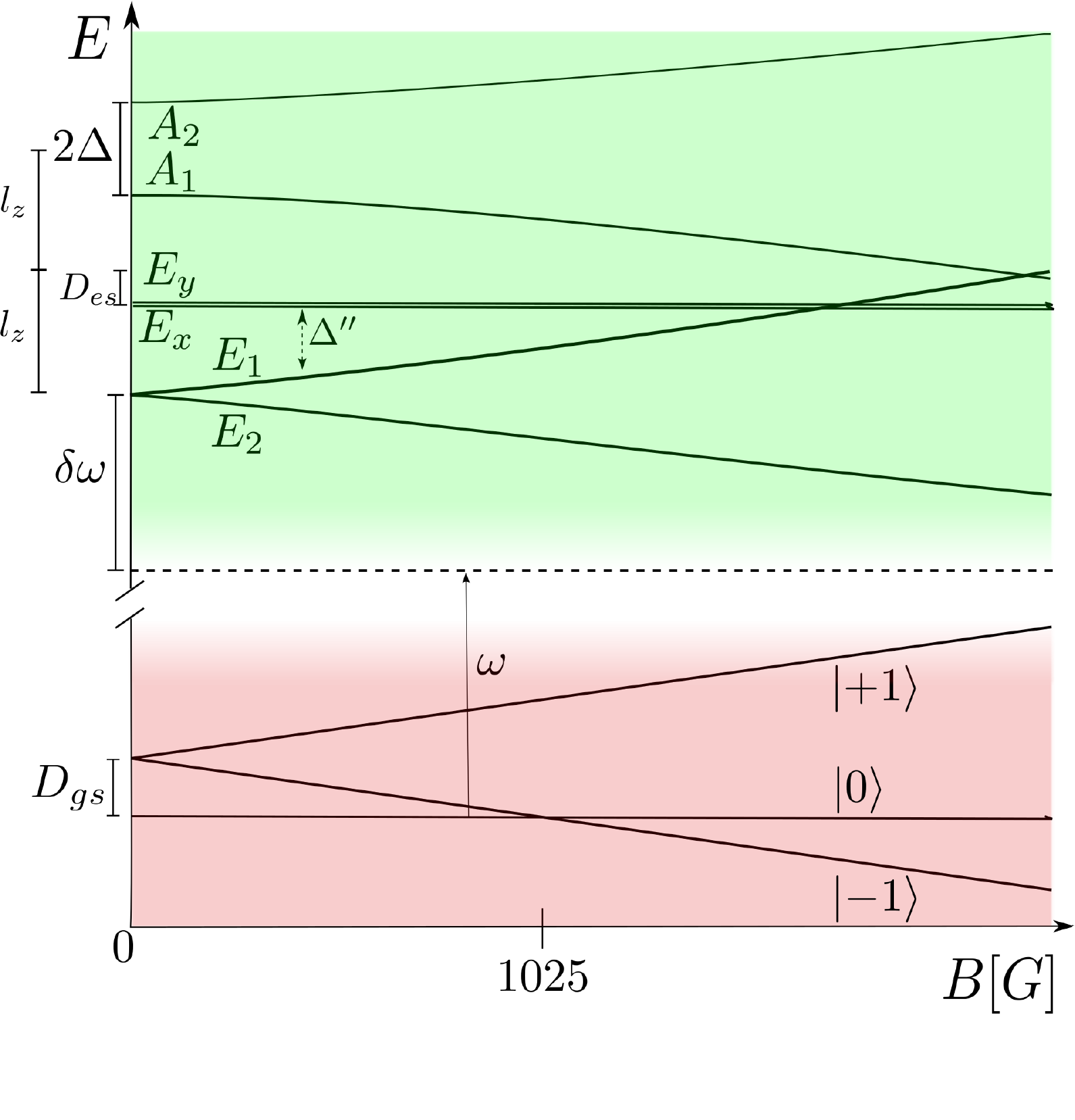}
\caption{Illustration of the optical pumping process with frequency
  $\omega$ from $m_s = 0$ 
ground state spin level below the excited state. 
The degenerate $E_{x,y}$ levels are coupled to the
$E_{1,2}$ levels by the transversal spin-spin-interaction $\Delta''$. \label{pumping}}
\end{figure}

At low temperatures, the excited state fine structure can be
understood to a large extent from strain and spin-spin interactions. 
In the basis of spin-orbit states with full symmetry, described by
the $C_{3v}$ double group including spin \cite{Maze}, 
$\left\{A_1, A_2, E_X, E_Y, E_1, E_2\right\}$
the excited-state Hamiltonian matrix is
\begin{widetext}
\begin{equation}\label{Hes}
H_{\rm es} = 
\left(\begin{matrix}
D_{\rm es}/3 + \Delta + l_z & g_{\rm es}\mu_B B & 0 & 0 & \delta_x & -i \delta_y 
\\
g_{\rm es}\mu_B B &  D_{\rm es}/3 - \Delta + l_z & 0 & 0 & i \delta_y & -\delta_x 
\\ 
0 & 0 & -2 D_{\rm es}/3 + \delta_x & \delta_y & 0 & \Delta''
\\
0 & 0 & \delta_y &  - 2 D_{\rm es}/3 - \delta_x & i\Delta'' & 0 
\\
\delta_x & -i \delta_y & 0 & -i\Delta'' &  D_{\rm es} / 3 - l_z  & -g_{\rm es}\mu_B B
\\
i\delta_y & -\delta_x & \Delta'' & 0 & -g_{\rm es}\mu_B B &  D_{\rm es}/3 - l_z 
\end{matrix}
\right),
\end{equation}
\end{widetext}
where $l_z$ is the axial spin-orbit splitting, and
$D_{\rm es} = -\frac{3}{4} D_{zz}$ and $\Delta = \frac{1}{2}D_{x^2-y^2}$
are the well-known spin-spin interactions \cite{Len}. 
Experimentally, it was found that 
 $l_z = 5.3$ GHz, $D_{\rm es} = 1.42$ GHz and $\Delta = 1.55$ GHz \cite{Bat}. 
The Land\'{e} factors of ground and excited state were found to be equal
\cite{Fuc,Neu,Hem}, 
$g_{\rm gs} \simeq g_{\rm es}   \simeq 2.01  = g$.
The energy gap $E_g$ is defined as the difference between the
$E_{1,2}$ excited states and the $m_S=\pm 1$ ground states at $B=0$.
The dependence of the ground- and excited state levels on $B$ is
shown in Fig.~\ref{pumping}.

Since electric dipole transitions are spin-conserving, our all-optical
spin control scheme requires a spin non-conserving mechanism in the excited state.
The longitudinal spin-orbit interaction term $l_z$ only leads to an
additional energy splitting between states with different spin
projections and cannot flip the spin.  
It was speculated that the transversal part of the spin-orbit 
interaction can lead to spin flips  \cite{Rog, Tam} , but it has
recently turned out that it can only connect orbital states belonging to different irreducible 
representations\cite{Maze}.
However, the transversal component $\Delta ''\simeq 0.2\,{\rm GHz}$ of
the spin-spin interaction allows for the non-spin-conserving 
transitions between the $E$ and $E'$ states, explaining the experimentally
observed transitions\cite{Maze}.

The components $\delta_x$ and $\delta_y$ of the non-axial strain 
can be written in polar coordinates,
$\delta_\bot^2 = \delta_x^2 + \delta_y^2$ and 
$2\beta = \arctan(\delta_y/\delta_x)$.  
Here, $\beta$ was defined such that it corresponds to  
the angle between the symmetry axis of strain 
eigenstates $e_x'\left(\delta_x,\delta_y\right)$ and the symmetry 
axis of the unperturbed $e_x$-orbital.

\subsection{Electric dipole transitions}

We assume the system to be optically driven with a radiation field at
fixed frequency $\omega$ near $E_g$, 
therefore it is convenient to describe the excited states
in a corotating frame, while keeping the ground states
fixed.  We then work in  the rotating wave approximation
where counter-rotating terms with frequency $E_g + \omega$ are
neglected.  This is justified as long as $\delta \omega =E_g - \omega
\ll E_g$.
Optical transitions between the ground and excited state are described
with the electric dipole operator for two electrons,
$H_{\rm dip}^{(2)} = H_{\rm dip}^{(1)}\otimes\mathbb{1} + \mathbb{1} \otimes  H_{\rm dip}^{(1)}$,
where 
\begin{equation}
 H_{\rm dip}^{(1)} = e \mathbf{E}\cdot\hat{\mathbf{r}} =
 e\left|\mathbf{E}\right|\left( \hat{x}\cos{\alpha} + \hat{y}\sin{\alpha}\right)
\end{equation}
is the single-particle electric dipole operator, where $\hat{\mathbf{r}} =
(\hat{x},\hat{y},\hat{z})$
denotes the electron position operator. 
Here, we assumed the incident light to be linearly polarized perpendicular to the $NV$-axis 
(which defines the $z$-axis of the coordinate system) with polarization angle $\alpha$, with $\alpha = 0$ for polarization parallel to the symmetry axis 
of the $e_x$-orbital.
Taking matrix elements with the ground and excited state basis
states yields the Hamiltonian
\begin{equation}\label{ham} 
H = \begin{pmatrix} 
               H_{\rm gs} & 0 \\ 0 &  \Delta\omega \mathbb{1} + H_{\rm es}
     \end{pmatrix}
 + 
\begin{pmatrix}
 0 & v \\ v^\dagger & 0
\end{pmatrix} 
= H_0 + V,
\end{equation}
where the detuning $\delta\omega$ is defined with respect to the 
lowest-lying excited state energy levels $E_{1,2}$ (neglecting strain and spin
mixing $\Delta''$), and $\Delta\omega = \delta\omega + l_z - \frac{D_{\rm es}}{3}$.
The transition matrix is given as
\begin{equation}
v = \begin{pmatrix}
i\epsilon_+ & -i\epsilon_+ & 0 & 0 & -i\epsilon_- & -i\epsilon_-
\\
0 & 0 & -2\epsilon_y & 2\epsilon_x & 0 & 0 
\\
-i\epsilon_- & -i\epsilon_- & 0 & 0 & i\epsilon_+ & -i\epsilon_+ \end{pmatrix}
\end{equation}
where, for linear polarization of the excitation field,
\begin{equation}
\epsilon_\pm  = \epsilon_x \pm i\epsilon_y \equiv \epsilon e^{\pm i\alpha},
\end{equation}
and ($i=x,y$)
\begin{equation}
\epsilon_{i} = \frac{\left\langle e\right|\left|\hat{r}_E\right|\left|a\right\rangle}{4} e E _{i}
\end{equation}
with the reduced matrix element of the position operator defined as
\begin{equation}
 \left\langle e\right|\left|\hat{r}_E\right|\left|a\right\rangle = \left\langle e_x\right|\hat{x}\left|a\right\rangle = 
\left\langle e_y\right|\hat{y}\left|a\right\rangle .
\end{equation}
We are interested in linearly polarized optical fields, where
$\epsilon$ is real.
The magnitude of the dipole matrix elements can be estimated from 
the observed Rabi oscillation period \cite{Rob} and the linear Stark shift \cite{Tam2},
typically $\epsilon\approx 1 \,{\rm GHz}$.

\section{Effective Spin Hamiltonian}

\subsection{Schrieffer-Wolff transformation}

Since the energy levels of the ground state $H_{\rm gs}$ and the 
excited state $H_{\rm es}$ are widely seperated by 
$E_g = 1.945$ eV $\approx 470$ THz and coupled by small perturbations
$\epsilon \ll |\delta\omega| = |E_g-\omega|$, we can use a
Schrieffer-Wolff transformation \cite{SW} of 
the Hamiltonian (block-matrix) as a valid approach to determine the effective dynamics of the driven system up to
second order in the perturbation. 
The Schrieffer-Wolff transformation is defined as follows,
\begin{equation}\label{Heff}
\tilde{H} = e^{S}He^{-S} = H + \left[S,H\right] + \frac{1}{2}\left[S,\left[S,H]\right]\right] + \mathcal{O}\left(S^3\right),
\end{equation}
with the anti-hermitian transformation matrix
\begin{equation}
 S = \begin{pmatrix}0 & s \\ -s^\dagger & 0 \end{pmatrix} .
\end{equation}
The aim of the transformation is to remove the coupling $V$ in first
order, which can be achieved if $\left[S,H_0\right]=  -V$.
In terms of the submatrices for the two seperated systems $H_{\rm gs}$ and $H_{\rm es}$ and their coupling $V$ this condition reduces to
\begin{equation}
s H_{\rm es} - H_{\rm gs}s = -v,
\end{equation}
which also implies that in a perturbation series in $v$, the matrix
$s$ will be first order, $s \sim O(v)$.
This particular choice of transformation secures that the first order terms in Eq.~(\ref{Heff}) cancel and we are left with an effective ground-state Hamiltonian
\begin{equation}
\tilde{H} = H_{\rm gs} + \frac{1}{2}\left(s^\dagger v + v^\dagger s\right) + \mathcal{O}\left(v^3\right).
\end{equation}
Note that in the absence of strain $l_z - D_{\rm es} - D_{\rm gs} \approx 1$
GHz is the seperation between the two closest-lying energy levels
$E_{x,y}$ and $E_+ = E_1 + E_2$, and thus
for resonant excitation between those two levels, the optical driving field strength has to be much smaller 
than $500$ MHz.

\subsection{Rotation axis}

\begin{figure}[t]
\includegraphics[width = 0.3\textwidth]{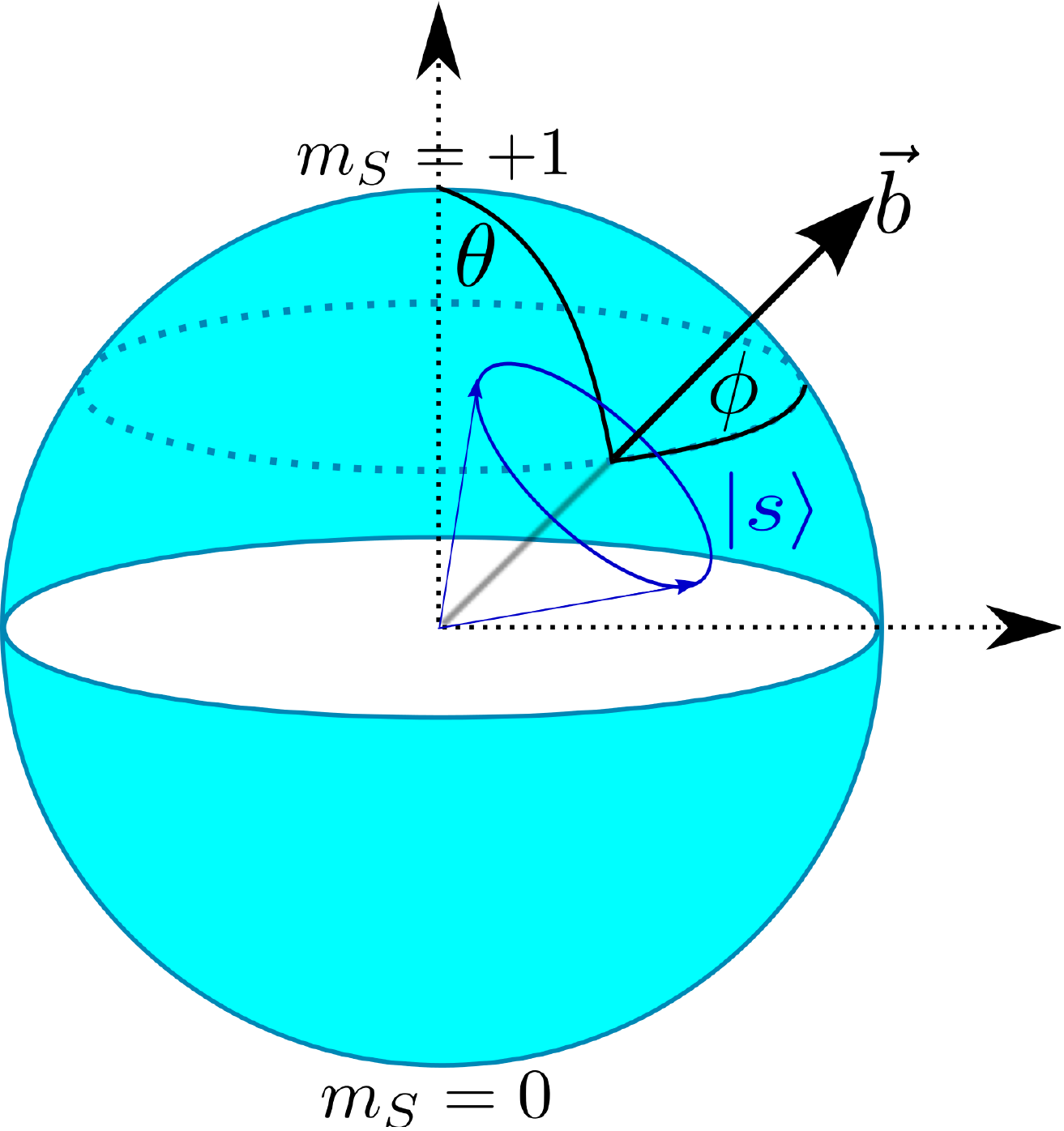}
\caption{The state $|s\rangle$ of the ground state spin qubit in the
  $m_S=0,-1$ subspace can be represented as a vector on the
  Bloch sphere where the two poles represent the two Zeeman-split eigenstates 
$\left|m_s = 0\right\rangle$ and $\left|m_s = -1\right\rangle$.
The vector $\mathbf{b}$ denotes the effective magnetic field acting on
this pseudo-spin 1/2, with spherical angles $\theta$ and $\phi$.
\label{blochsphere}}
\end{figure}
We focus on the transition between the $m_S = -1$ and $m_S = 0$ ground
state spin levels.
This two-level system can be split off from the $m_S=+1$ state \cite{footnote_SW2}, 
in the regime $g\mu_B |\delta B|  \lesssim 2D_{\rm gs}$.
In this case the dynamics is described by the $2\times 2$ Hamiltonian
\begin{equation}
\tilde{H} = 
\left(\begin{matrix}
\tilde{H}_{11} & \tilde{H}_{12}
\\
\tilde{H}_{21} & \tilde{H}_{22}
\end{matrix}\right)
= b_0\openone + \bm{\mathbf{b}} \cdot \bm{\sigma}
\label{eq:qubitH}
\end{equation}
where the effective (pseudo-) magnetic field has components,
\begin{eqnarray}
b_x &=& \frac{1}{2}\left(\tilde{H}_{12}+\tilde{H}_{21}\right) = b_\bot \cos\phi,\\
b_y &=& \frac{1}{2i}\left(\tilde{H}_{12}-\tilde{H}_{21}\right) = b_\bot \sin\phi,\\
b_z &=& \frac{1}{2}\left(\tilde{H}_{11}-\tilde{H}_{22}\right),
\end{eqnarray}
and $b_0= \tilde{H}_{11}+\tilde{H}_{22}$.
The dynamics within this two-dimensional subspace can be visualized 
using a Bloch sphere picture, as shown in Fig.~\ref{blochsphere}.
Optical driving results in a rotation of the state vector about an axis $\mathbf{b}$
with polar angle (axial orientation) $\theta $ and azimuthal angle (non-axial orientation) $\phi$. 
The transversal part $b_\bot = \left|\tilde{H}_{12}\right|=\sqrt{b_x^2+b_y^2}=b\sin\theta$ of
the effective field provides for
effective spin-flip transitions while $b_z$ accounts 
for the effective (AC Stark) splitting 
between the two levels (around the LAC). The term $\propto
\mathbbm{1}$ in Eq.~(\ref{eq:qubitH})
can be omitted 
since it merely leads to a global phase. 
The precession frequency is given by 
\begin{equation}
 b = \sqrt{b_\bot^2 + b_z^2} = \sqrt{1 +
   \cot^2{\theta}}  |b_\bot |.
\end{equation}
Spin flips can be implemented as rotation about an axis within the
equatorial plane of the Bloch sphere (Fig.~\ref{blochsphere}) which 
corresponds to $\theta = \frac{\pi}{2}$ and thus $b = b_\bot$.

\section{Results}

\subsection{Unstrained NV center}\label{zerostrain}
In the case of vanishing strain ($\delta_\bot = 0$) we obtain a simple
analytical result for the transversal component of the qubit rotation axis,
with magnitude,
 \begin{eqnarray}
  {b}_\bot &=&
  \Delta''\epsilon^2\left|\frac{1}{\delta\omega\left(\delta\omega +
        D_{\rm es} + l_z + D_{\rm gs} + g\mu_B\delta B
      \right)}\right.   \label{inplane-axis}\\
& & \!\! + \left.\frac{1}{\left(\delta\omega-g\mu_B\delta B\right)\left(\delta\omega - D_{\rm es} + l_z + D_{\rm gs} \right)}
\right| + O\!\left(\Delta''^2\right)\!,\nonumber
 \end{eqnarray}
which is proportional to the intensity $\epsilon^2$ of the optical
driving field and the transversal spin-spin coupling $\Delta''$ in the
excited state.
The azimuthal angle of the rotation axis is determined 
by the optical polarization angle $\alpha$,
\begin{align}\label{phizero}
\phi = - 2\alpha
\end{align}
where the factor of 2 reflects the double group character of spin representation.
The polar angle $\theta$ of the rotation axis is independent of
$\alpha$ and for small $g\mu_B\delta B$ even independent of the driving field strength $\epsilon$.
The residual Zeeman splitting is limited by the hyperfine LAC of about
$2$ MHz\cite{Gali}, and therefore for an optical coupling $\epsilon
\gtrsim 100\,{\rm MHz}$ we 
can always find pairs of parameters
 $(\delta B,\delta \omega)$ that fullfill the condition $\theta=\pi/2$.

In the limit of large detuning, i.e., when $\delta \omega$ dominates
all other energies in the denominators of Eq.~(\ref{inplane-axis}), we
can approximate the transverse component of the effective field as
\begin{equation}
b_\bot \simeq 2\Delta''\frac{\epsilon^2}{\delta \omega^2}.
\end{equation}

\subsection{Effect of strain}
\begin{figure}
\centering
 \includegraphics[width=0.4\textwidth]{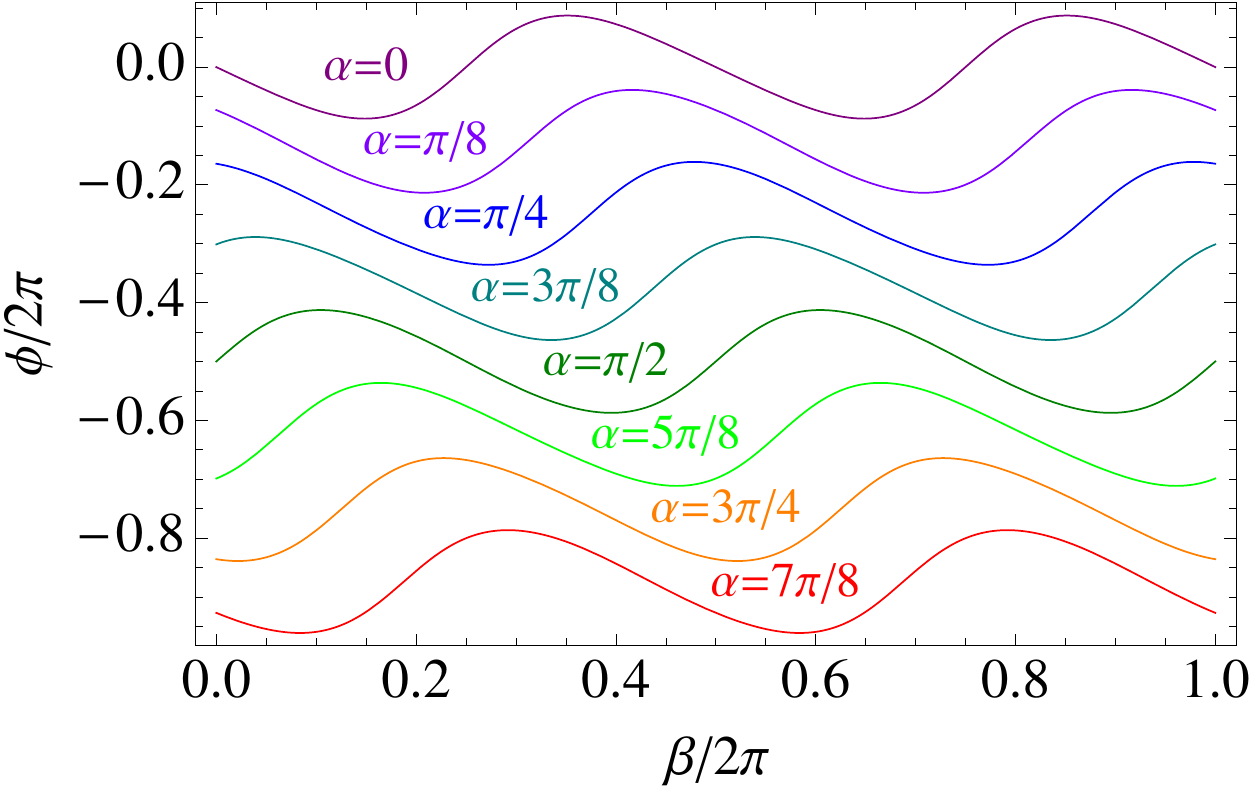}
\caption{Non axial orientation angle $\phi$ of precession axis for different polarization angles $\alpha$ in respect to strain orientation 
at transversal strain $\delta_\bot = 5\,{\rm  GHz}$. \label{phialpha}}
\end{figure}

We now include the effect of strain in the diamond crystal into our discussion.
For moderate strain $\delta_{x,y} \ll 2l_z \approx 10$ GHz, the 
$A_{1,2}$ levels of the excited state are largely seperated from the
$E_{1,2}$ levels, and thus the strain-induced mixing of $A_{1,2}$ states and 
$E_{1,2}$ states can be neglected in lowest order. 
\begin{figure}[htb!]
\centering
  \includegraphics[width=0.435\textwidth]{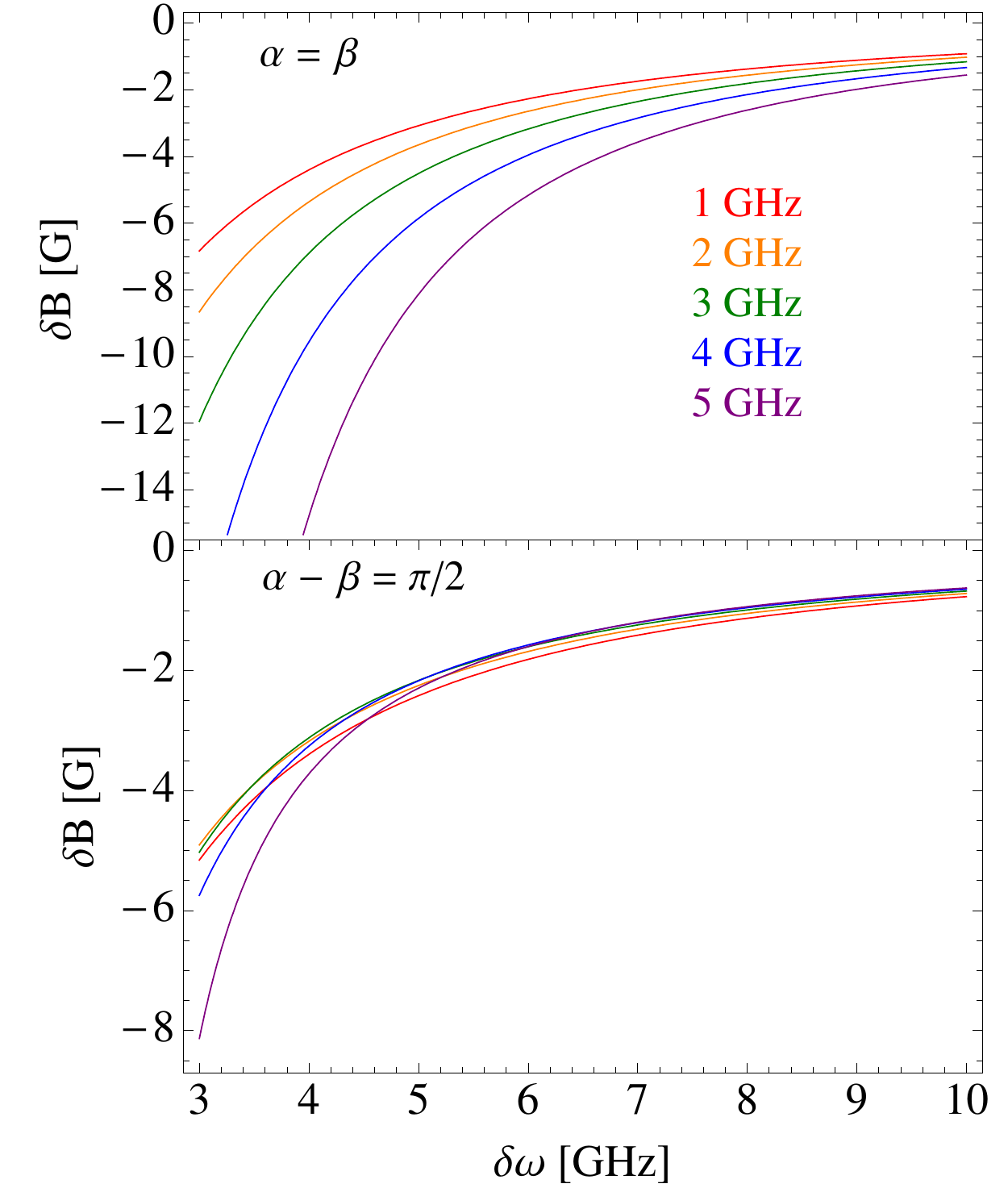}
\caption{Pairs of required magnetic field variation $\delta B$ and detuning $\delta\omega$ at fixed dipole coupling strength 
$\epsilon = 500 \,{\rm MHz}
$ for different values of transversal strain $\delta_\bot = 1, 2, 3, 4$ and $5$ GHz (from top to bottom) to fullfill the condition 
$\theta = \frac{\pi}{2}$ for precession around a rotation axis within
the equatorial plane of the Bloch sphere for $\alpha = \beta = 0$.
\label{deltaBrootstr}}
\end{figure}
\begin{figure}[htb!]
\centering
\includegraphics[width=0.435\textwidth]{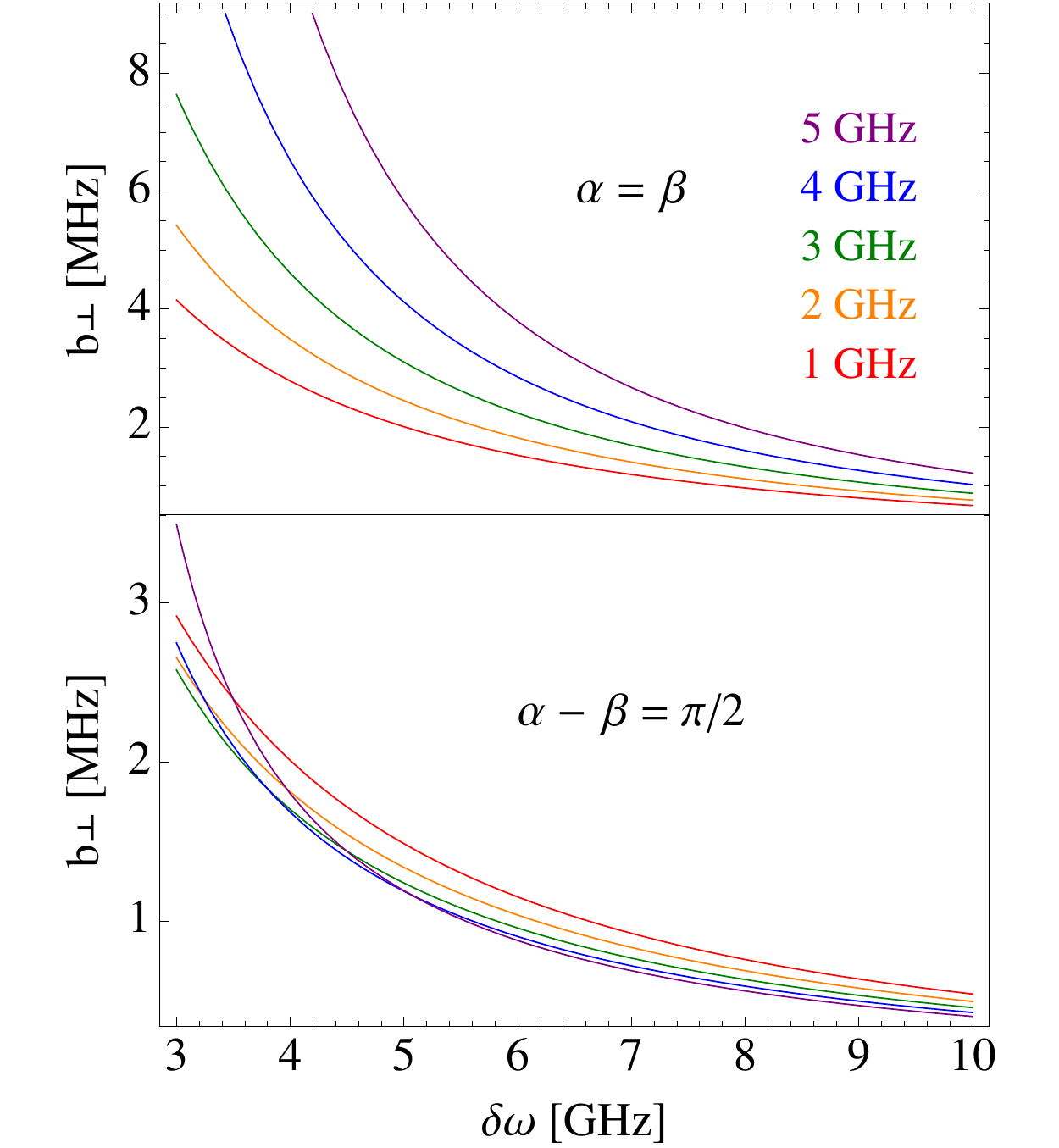}
\caption{Precession frequency $b_\bot$ for different values of
  transversal strain $\delta_\bot = 1, 2, 3, 4$ and $5$ GHz 
(from left to right) for 
polarization parallel ($\alpha = \beta$) and perpendicular ($\alpha - \beta = \frac{\pi}{2}$) to strain with dipole coupling $\epsilon = 500$ MHz.\label{Btrootstr}}
\end{figure}
The main effect of moderate strain is thus a shift of the
resonances in Eq.~(\ref{inplane-axis}) by $\pm \delta_\bot = \pm \sqrt
{\delta_x^2+\delta_y^2}$ , lifting the degeneracy of $E_x$ and $E_y$ levels. Though strain does 
not directly mix states with different spin projections, this shift
reduces the energetic seperation 
between coupled $E$ and $E'$ levels and therefore strongly enhances the efficiency of spin-flip-transitions.
\begin{figure}[htb!]
\centering
 \includegraphics[width=0.435\textwidth]{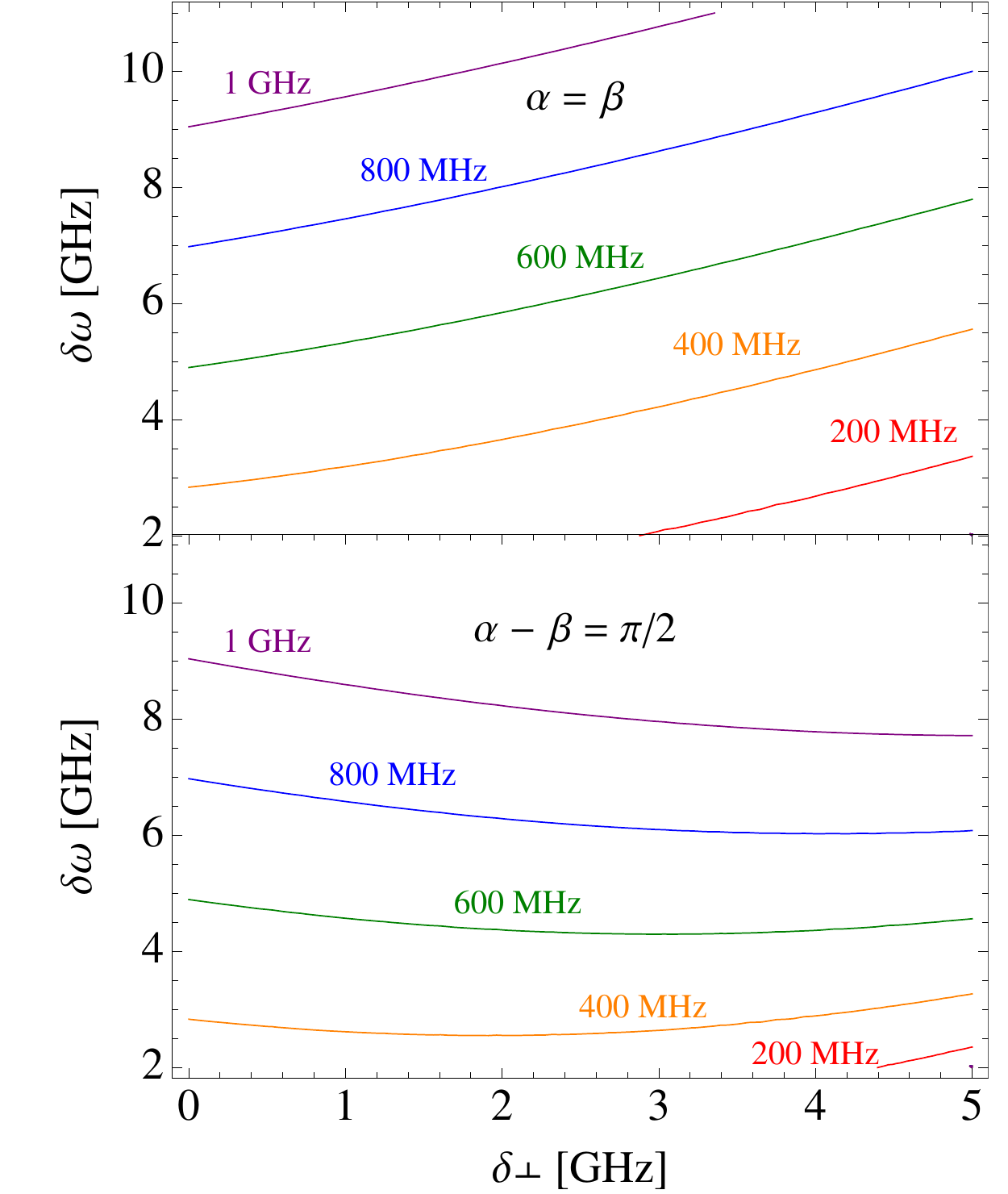}
\caption{Pairs of detuning $\delta\omega$ and transversal strain $\delta_\bot$ that match the condition $\theta = \frac{\pi}{2}$ for precession around a
rotation axis within the equatorial plane of the Bloch sphere for different dipole coupling
$\epsilon = 100, 200, 400, 600, 800$ and $1000$ MHz (from bottom to top) for $\alpha = \beta = 0$.}
\end{figure}
In Fig.~\ref{deltaBrootstr}, we plot suitable pairs of parameter
values for the detuning $\delta\omega$ (near resonant driving) and
Zeeman splitting $g\mu_B\delta B$ with varying strain $\delta_\bot$
that fulfill the condition $\theta = \frac{\pi}{2}$ for an in-plane rotation axis.
\begin{figure}[htb!]
 \includegraphics[width = 0.435\textwidth]{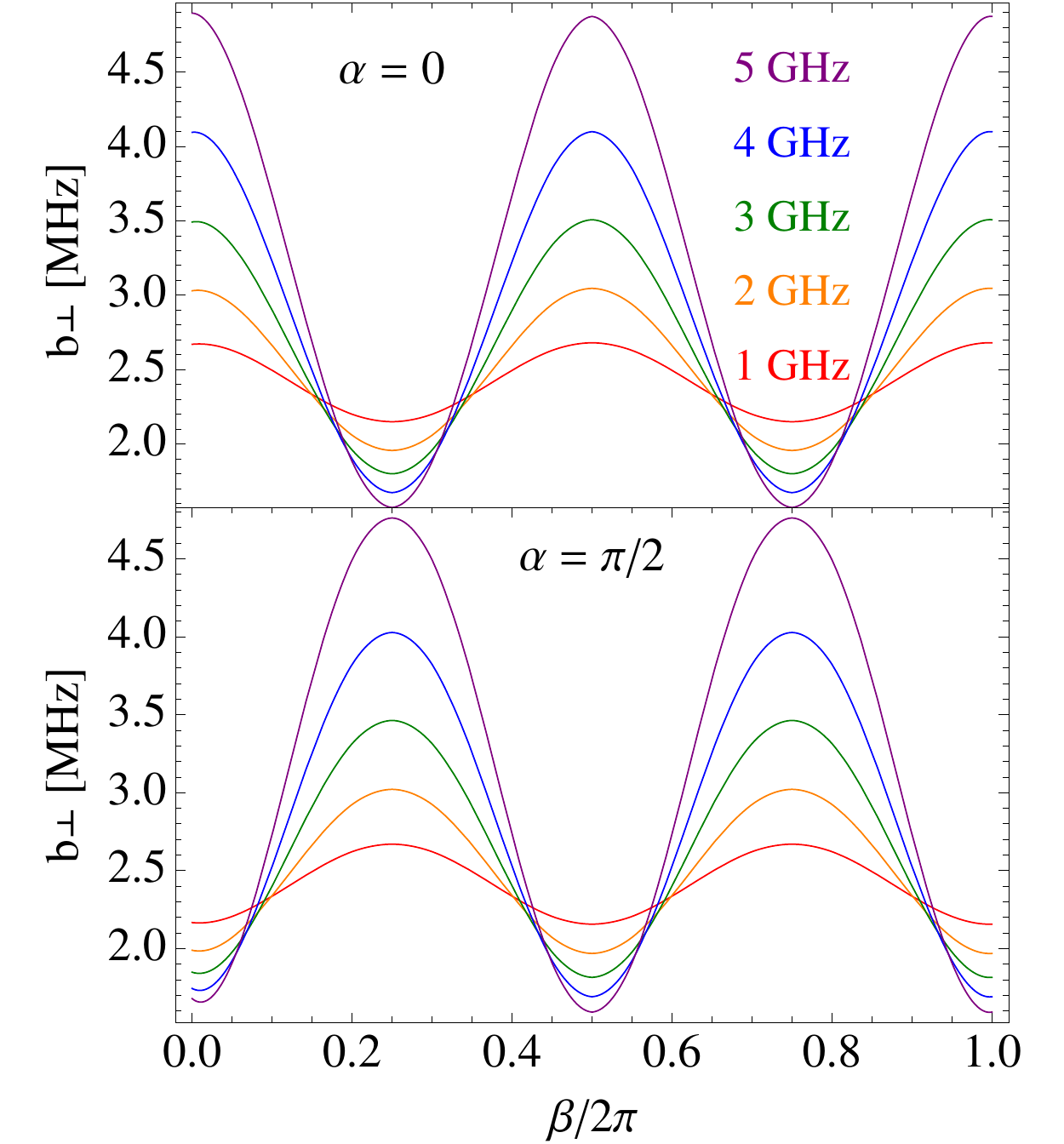}
\caption{Precession frequency $b_\bot$ around in-plane rotation axis at $10$ GHz detuning for optical coupling $\epsilon = 500$ MHz and polarization
 in $e_x$ ($\alpha = 0$) and $e_y$ ($\alpha = \frac{\pi}{2}$)
 direction for different values of transversal strain 
$\delta_\bot = 1,2,3,4,5$ GHz (increasing amplitude) over strain direction angle $\beta$.
\label{Btbeta}}
\end{figure}
This defines an implicit function $\delta B\left(\delta\omega, \delta_\bot\right)$
given $\alpha$ and $\beta$, e.g. in the case of $\theta=\pi/2$ for $\alpha = \beta = 0$. 
To find the strength of spin-flip transitions, we substitute 
$\delta B$ into the precession frequency for an in-plane rotation axis, $b_\bot$ 
and plot it in Fig.~\ref{Btrootstr} as a function of the optical
frequency detuning $\delta \omega$ for different values of transversal strain.
We find that the precession frequency $b_\bot$ indeed increases with strain.
Varying $\epsilon$ numerically shows that the precession frequency is
still proportional to the intensity of the optical driving field
$b_\bot|_{\theta = \frac{\pi}{2}} \propto \epsilon^2$.
Note that the perturbative approach breaks down as detuning approaches
strain (divergence of $b_\bot$ in Fig. ~\ref{Btrootstr}), restricted by the 
validity condition $\delta\omega \ll \delta_\bot$ 
for the Schrieffer-Wolff transformation in Eq.~(\ref{ham}).
We also investigate the dependence on the direction of strain and
polarization (see Fig. ~\ref{Btbeta}). Expectedly we get 
the highest efficiency for collinear 
strain and polarization and a minimal efficiency 
for perpendicular relative orientation with an overall sinusoidal form of twofold symmetry. Changing the optical polarization angle $\alpha$ only leads to a uniform and continuous shift of this function 
(this has been checked for a variety of different values, but for
simplicity we only show it for $\alpha = 0$ and $\alpha = \frac{\pi}{2}$). 
Thus, for weak strain, the resulting effective field 
only depends on the relative angle $\alpha - \beta$ between the strain and
polarization angles. 
However, as the transversal strain $\delta_\bot$ increases beyond
about $10$ and $20$ GHz, we start observing a
modulation of the field with higher harmonics of  $\alpha - \beta$.
\begin{figure}
\centering
 \includegraphics[width=0.4\textwidth]{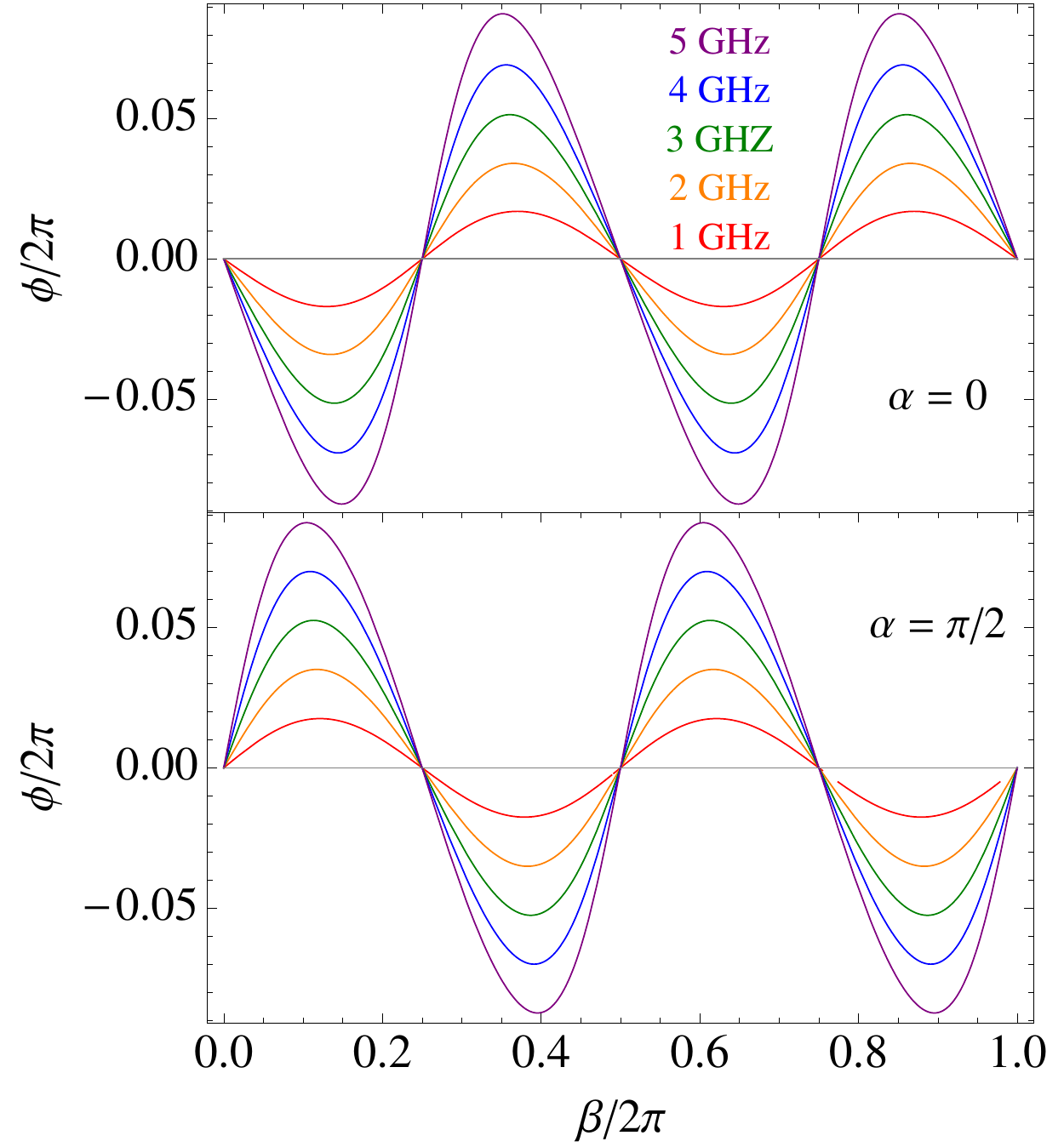}
\caption{Non axial orientation angle $\phi$ of precession axis for polarization parallel ($\alpha = 0$) and perpendicular ($\alpha = \frac{\pi}{2}$) to $e_x$-orbitals symmetry axis
 in respect to strain orientation 
for different strengths of transversal strain. \label{phibeta}}
\end{figure}
Numerical evaluation also reveals that the azimuthal angle (for
in-plane orientation of the precession axis) 
is generally independent of optical 
coupling strength and magnetic field,
\begin{equation}
 \phi = \phi\left(\alpha,\beta,\delta_\bot,\delta\omega\right).
\end{equation}
From Fig.~\ref{phibeta} we see that for $e_{x/y}$-polarized light
(\textit{i.e.} $\alpha = 0,\frac{\pi}{2}$), the
angle $\phi$ is well approximated by 
$\phi = \mp\hat{\phi}\left(\delta_\bot,\delta\omega\right)\left(\delta_\bot,\delta\omega\right)
\sin{2\beta}$ (at low strain), with an amplitude proportional to the
intensity of the strain,
\begin{equation}
 \hat{\phi} = \delta_\bot f\left(\delta\omega\right)
\end{equation}
In Fig.~\ref{phialpha}, we show that the 
sinusoidal shape of $\phi$ as a function of $\alpha$ is 
slightly distorted for polarization angles $\alpha \neq n\frac{\pi}{2}$, $n \in \mathbb{N}$;
we also find that this distortion grows with increasing strain.

\subsection{High strain limit}

In the high strain regime, the electronic states of the NV-center are 
energetically split into two orbital branches with largely seperated energies $E_x$ and $E_y$ 
corresponding to a specific choice of coordinate axes, that fixes the orientation angle $\beta$.
In this limit $b_\bot$ and $b_z$ (and thus the polar angle $\theta$)
become independent of the orientation angles
of both strain and polarization.
Expanding in $\delta_\bot^{-1}$ yields
\begin{equation}
 b_\bot \simeq 2\Delta''\frac{\epsilon^2}{\delta_\bot^2},
\end{equation}
and 
\begin{equation}
 \phi = 2\left(\alpha - 2\beta\right) - \frac{\pi}{2}+ \mathcal{O}\left(\frac{1}{\delta_\bot}\right).
\end{equation}
for the transversal part of the pseudo-field, and 
\begin{equation}
  b_z = -\frac{g\mu_B\delta B}{2} + 2\left(D_{\rm es}-D_{\rm gs}\right)\frac{\epsilon^2}{\delta_\bot^2} +  \mathcal{O}\left(\frac{1}{\delta_\bot^3}\right),
\end{equation}
for the longitudinal part.
The higher orders contain higher harmonics, such as terms $\propto\Delta \cos{2\left(\alpha - 3\beta\right)}$ for  $b_x$ and 
$\propto\Delta \sin{2\left(\alpha - 3\beta\right)}$ for  $b_y$
in $\mathcal{O}\left(\frac{1}{\delta_\bot^3}\right)$, 
indicating strain induced third order transitions mediated by the
$A_{1,2}$ levels.

\section{Conclusion}
We have shown that the effective precession axis and frequency of the
ground state spin of the $NV^-$-center can be fully controlled by
off-resonant optical excitation, by
adjusting the frequency detuning $\delta\omega$ and linear polarization angle
$\alpha$ of the optical driving field for a given intensity $\epsilon$ (optical
dipole coupling) and magnetic field $B$. 
The orientation of the precession axis is determined by two angles $\theta$ and $\phi$, 
where the first is depending on all parameters (including strain and
polarization) and the 
latter is independent of magnetic field and 
optical coupling strength and basically controlled by polarization and
strain. 
The strain effects can be compensated by external bias voltage \cite{Bassett2011}.
Since any unitary qubit operation (rotation around axis $\mathbf{n}$ by
angle $\gamma$) can be composed by 
successive rotations around two orthogonal axes on the 
Bloch sphere, a complete set of single-qubit operations can be
generated optically in this way.
From a purely geometric point of view, spin rotation about an axis within the equatorial plane of the Bloch sphere 
(where $b_z = 0$) is most effective 
for flipping the spin, although any axis other than the $z$-axis
would do (the smaller the polar angle $\theta$, the more pulses 
are be required).  
A full spin-flip is obtained by a
$b_\bot \tau = \pi$-rotation around an axis within the equatorial 
$(x,y)$-plane of the Bloch sphere,
providing an estimate for the gate switching (optical pumping) time
$\tau$ in the limit of large detuning and weak strain,
\begin{equation}
\tau \simeq \frac{\pi}{2}\frac{\delta\omega^2}{\Delta''\epsilon^2}.
\end{equation}
The switching time for spin-flip transitions is limited by the spin
mixing term $\Delta''$, since the (above) condition implies
via the off-resonant condition $\epsilon\ll\delta \omega$
the following lower limit for the spin-flip,
\begin{equation}
 \tau \gtrsim \frac{\pi}{\Delta''}\approx 10\,\mathrm {ns}
\end{equation}
where for that latter estimate we assumed $\Delta'' = 0.2$ GHz.

\section*{Acknowledgments}
We thank B. Buckley for discussions and
F. Fehse for generating the graphics in Fig.~\ref{NV}
and the Konstanz Center for Applied Photonics (CAP) and BMBF
QuHLRep for funding.


\begin{thebibliography}{99}
\bibitem{footnote_SW2}
This can be justified by a second Schrieffer-Wolff transformation.
\end{thebibliography}
\end{document}